\documentclass[mathleft]{an}
\usepackage{graphicx}
\usepackage{times}
\renewcommand{\vec}[1]{\mbox{\boldmath$#1$}}
\def\gsim{\lower.4ex\hbox{$\;\buildrel >\over{\scriptstyle\sim}\;$}}
\def\lsim{\lower.4ex\hbox{$\;\buildrel <\over{\scriptstyle\sim}\;$}}
\Yearsubmission{2011}
\Month{11}
\begin{document}

\title{Suppression  of the large-scale Lorentz force by turbulence}
\titlerunning{Suppression  of the large-scale Lorentz force by turbulence}
\authorrunning{ G. R\"udiger, L.L. Kitchatinov \& M. Schultz}
\author{G.~R\"udiger\inst{1}\fnmsep\thanks{Corresponding author: gruediger@aip.de}  \and L.L. Kitchatinov\inst{2,3} \and M. Schultz\inst{1}}
\institute{Leibniz-Institut f\"ur Astrophysik Potsdam, An der Sternwarte
             16, D-14482 Potsdam, Germany
\and Institute for Solar-Terrestrial Physics, P.O.~Box 291,
      Irkutsk 664033, Russia\and
 Pulkovo Astronomical Observatory, St. Petersburg 196140,
 Russia}

\received{2011 Nov 17} \accepted{2011 Dec 5} \publonline{2012 Jan 23}

 \abstract{%
The components of the total  stress tensor (Reynolds stress plus Maxwell stress) 
  are  computed within  the quasilinear approximation for a driven turbulence  influenced by 
  a large-scale magnetic background field. The conducting fluid has an 
arbitrary magnetic Prandtl number and the turbulence without the  background field 
is assumed as homogeneous and isotropic  with a free  Strouhal number ${\rm St}$. 
The total large-scale  magnetic tension  is always reduced by the turbulence  with 
the possibility of a `catastrophic quenching' for large  magnetic Reynolds number
 $\rm Rm$ so that even its sign is reversed.   The total magnetic pressure is 
 enhanced by turbulence in the high-conductivity limit  
  but it is reduced in the low-conductivity limit. Also in this case the sign of 
the total pressure may reverse but only for  special turbulences with sufficiently large ${\rm St> 1}$. 
 The turbulence-induced terms of the  stress tensor are suppressed by  strong magnetic fields.
 For the tension term this quenching grows with the square of the Hartmann  number of 
the magnetic field. For microscopic  (i.e.  small) diffusivity values 
 the magnetic tension term becomes thus highly quenched even for field amplitudes
 much smaller than their equipartition value. In the opposite case of large-eddy 
simulations the magnetic quenching is only mild  but then also the turbulence-induced
 Maxwell tensor components for weak fields remain rather small.
}


\maketitle

\section{Introduction}\label{introduction}
Differential rotation and fossil fields do not coexist. A nonuniform rotation law
 induces azimuthal fields $\delta B_\phi$ from an  original poloidal field $B_R$ which
 together transport angular momentum in radial direction reducing the shear
 $\delta \Omega$ via the large-scale Lorentz force ${\vec J}\times {\vec B}$, i.e.
\begin{equation}
  R \frac{\delta\Omega}{ \delta t}
  \simeq \frac{B_R \delta B_\phi}{\mu_0\rho\Delta R}. 
  \label{-1}
\end{equation}
As the induced $B_\phi$ results as $\delta B_\phi\simeq \Delta \Omega B_R \delta t$ 
the duration of the complete decay of the shear (i.e. $\delta \Omega=\Delta \Omega$) is
$
  \delta t\simeq \sqrt{\mu_0\rho} R/B_R.  $ This is a short time of order 10\,000 yr for
 a fossil field of 1 Gauss compared with the time scale of the star formation. 
All protostars should thus rotate rigidly.

Equation (\ref{-1}) is also used for the explanation of the observed torsional oscillations 
of the Sun. With $B_R\simeq 5$~Gauss and $B_\phi\simeq 10\,000$~Gauss the estimation
 for $R\delta\Omega$ is 10 m/s which is close to  the observed value of 5 m\,s$^{-1}$. 
The result is that -- if Eq. (\ref{-1}) is correct -- the maximal field strength of
 the invisible toroidal fields should not be much higher than  10\,000~Gauss. 
 However, the solar convection zone is turbulent and it is not yet clear
  whether Eq. (\ref{-1}) is also true for conducting fluids with fluctuating flows and fields.

In this paper the total Maxwell stress is thus derived for a turbulent fluid under the presence of a uniform background field $\vec B$. The fluctuating flow components  are denoted by $\vec{u}$ and the fluctuating field components  are denoted by $\vec{b}$. The  standard Maxwell tensor
\begin{equation}
  M_{ij}= \frac{1}{\mu_0} B_iB_j - \frac{1}{2\mu_0}\vec{B}^2 \delta_{ij}
  \label{2}
\end{equation}
for the considered MHD turbulence turns into the generalized stress tensor
\begin{equation}
  M_{ij}^{\rm tot}= M_{ij} -\rho Q_{ij} +M_{ij}^{\rm T},
  \label{3}
\end{equation}
with the one-point correlation tensor 
\begin{equation}
Q_{ij}= \langle u_i(\vec{x},t) u_j(\vec {x},t)\rangle
 \label{33}
\end{equation}
of the flow
and the turbulence-induced  Maxwell tensor
\begin{equation}
 M_{ij}^{\rm T}= \frac{1}{\mu_0} \langle b_i(\vec{x},t)b_j(\vec{x},t)\rangle - \frac{1}{2 \mu_0}
\langle \vec{b}^2(\vec{x},t)\rangle \delta_{ij}.
  \label{3b}
\end{equation}
The generalized Lorentz force $\vec F$ is then
\begin{equation}
 F_i= M_{ij,j}^{\rm tot}.
  \label{3c}
\end{equation}
If the only preferred direction in the turbulence is the uniform background field $\vec B$  both the tensors $Q_{ij}$ and $M_{ij}^{\rm T}$ have the same form as the Maxwell tensor (\ref{2})  but with two unknown scalar parameters. It makes thus sense  to write
\begin{equation}
  M_{ij}^\mathrm{tot} = \frac{1}{\mu_0}(1-\kappa)B_iB_j - \frac{1}{2\mu_0}(1-\kappa_\mathrm{p})\vec{B}^2 \delta_{ij}
  \label{4}
\end{equation}
for the total stress tensor (\ref{3}). The first term of the RHS describes a tension 
along the magnetic field lines while the second term  is the sum of the magnetic-induced
 pressures transverse to the lines of force. The main role of the first term in
 stellar physics is an outward-directed angular momentum transport if  ${B_R \, B_\phi < 0}$. 
If its  coefficient ${1-\kappa}$ would  change its sign under the presence of turbulence then 
for the same magnetic geometry the angular momentum transport would be inwardly directed. 
The Lorentz force (\ref{3c}) with (\ref{4}) becomes
\begin{equation}
 \vec{F}=  (1-\kappa) \ \vec{J}\times\vec{B}  - \frac{1}{2\mu_0}(\kappa-\kappa_\mathrm{p})\ \nabla\vec{B}^2,
  \label{4b}
\end{equation}
so that the `laminar' Lorentz force $\vec{J}\times\vec{B}$ has to be multiplied with the 
factor ${1-\kappa}$ and an extra magnetic pressure  appears if the $\kappa$'s are unequal (and they are)
  due to the action of the turbulence. If the $\kappa$ is positive  then its 
  amplitude should
 not exceed unity as otherwise the direction of the Lorentz force reversed.
 Roberts \& Soward (1975) considering only the terms of the Maxwell stress found (large) positive $\kappa$ and 
 negative  $\kappa_\mathrm{p}$, i.e. 
 $\kappa=-\kappa_\mathrm{p}=\eta_{\rm T}/\eta$ with $\eta_{\rm T}$ the well-known eddy diffusivity (see Eq.
 (\ref{eta}),
 below).

R\"udiger et al. (1986)  found   $\kappa$ also as positive and as running with 
the magnetic Reynolds number $\rm Rm$ of the turbulence even 
for  $\rm Rm > 1$. Kleeorin et al. (1989) suggest 
that  $\kappa_\mathrm{p}>0$ and even larger than unity  
so that the total  magnetic pressure changes its sign and  
becomes negative. The resulting instability may produce 
structures of concentrated magnetic field and may be 
important for sunspot formation (Kleeorin et al. 1990; Brandenburg et al. 2010,
2011).

Hence, the $\kappa$'s have an important physical meaning. 
In the  simplest case they both would result as negative. 
Then  the  effective pressure is increased by the magnetic 
terms and also the tension term $1-\kappa$ remains positive 
so that the Lorentz force in turbulent media is simply amplified. 
Many more serious  consequences would result from positive $\kappa$ 
and $\kappa_\mathrm{p}$ if exceeding unity. In this case the Lorentz 
force changes its   sign with dramatic consequences for the theory of 
torsional oscillations and solar 
oscillations (Kleeorin \& Rogachevskii 1994: Kleeorin et al. 1996).

The turbulence-induced   modification of the Maxwell stress can 
also be important in other  constellations  where large-scale 
fields and turbulence simultaneously exist. Tachocline theory, 
jet theory, the structure of magnetized galactic 
disks (Battaner \& Florido 1995) or oscillations of 
convective stars with magnetic fields could be mentioned.

\section{Equations}
We apply the quasilinear approximation known also as the second order correlation approximation (SOCA). All preliminary steps to derive main equations for the problem at hand were described by R\"udiger \& Kitchatinov (1990). The fluctuating magnetic and velocity fields are related by the equation
\begin{equation}
  \hat{\vec b}({\vec k},\omega ) = \frac{\mathrm{i}({\vec k}\cdot{\vec B})}
  {- \mathrm{i}\omega+\eta k^2 } \hat{\vec u}({\vec k},\omega ) ,
  \label{5}
\end{equation}
where the hat notation marks Fourier amplitudes, e.g.
\begin{equation}
  {\vec b}({\vec r},t) = \int\mathrm{e}^{\rm{i}({\vec {k x}} - \omega t)}
  \hat{\vec b}({\vec k},\omega )\ \mathrm{d}{\vec k}\mathrm{d}\omega ,
  \label{6}
\end{equation}
and the same for the velocity field. The influence of mean magnetic field on turbulence is described by the relation
\begin{equation}
  \hat{\vec u}({\vec k},\omega ) =
  \frac{\hat{\vec u}^{(0)}({\vec k},\omega )}
  {1 + \frac{({\vec k}\cdot{\vec V})^2}
  {(-\mathrm{i}\omega +\eta k^2 )(- \mathrm{i}\omega+\nu k^2 )}},
  \label{7}
\end{equation}
where ${\vec V} = {\vec B}/\sqrt{\mu_0\rho}$ is the  Alfv\'en velocity of the large-scale background field,
 $\eta$ is the microscopic magnetic diffusivity, and $\nu$ is the microscopic viscosity. 
In Eq.~(\ref{7}) $\hat{\vec u}$ is the (Fourier-transformed) velocity field modified by
 the mean magnetic field and $\hat{\vec u}^{(0)}$ stands for the velocity of
 the \lq original' turbulence which is assumed to exist for   $\vec{B} = 0$. 
 The original turbulence is assumed as statistically homogeneous and isotropic, i.e.
\begin{eqnarray}
\lefteqn{\langle{\hat{u}_i^{(0)}({\vec k},\omega)
  \hat{u}_j^{(0)}({\vec k}',{\omega}')\rangle} =
  \frac{E(k,\omega)}{16\pi k^2}\left(\delta_{ij}
  - \frac{k_ik_j}{k^2}\right)}
  \nonumber \\
  &&\ \ \ \ \ \ \ \ \ \ \ \ \ \ \ \ \ \qquad \qquad \qquad \times\ \delta ({\vec k} + \vec{k}')
  \delta (\omega + {\omega}') ,
  \label{8}
\end{eqnarray}
where $E(k,\omega )$ is the positive-definite spectrum function of the turbulence. Here
\begin{equation}
  {\vec u}^2 = \int\limits_0^\infty \int\limits_0^\infty\
  E(k,\omega )\ \mathrm{d}k \mathrm{d}\omega
\label{8.1}
 \end{equation}
defines the   rms velocity  $\vec{u}$ of the original turbulence. Equations (\ref{5}) to (\ref{8}) suffice to derive the values of the $\kappa$'s.

\section{Weak field}
 We proceed by considering  special cases. For weak  mean magnetic 
 field  one  finds from the expressions given 
 by R\"udiger \& Kitchatinov (1990) the relation
\begin{eqnarray}
\lefteqn{  \kappa = \frac{1}{15} \int\limits_0^\infty \!\! \int\limits_0^\infty
  \frac{E k^2 (\nu (2\eta + \nu )k^4 - \omega^2)}
  {(\nu^2k^4 + \omega^2)(\eta^2k^4 + \omega^2)}\
  \mathrm{d} k \mathrm{d}\omega ,}
  \nonumber \\
\lefteqn{  \kappa_\mathrm{p} = \frac{1}{15}
  \int\limits_0^\infty\!\! \int\limits_0^\infty
  \frac{E k^2 (\nu (8\eta - \nu )k^4 - 9\omega^2)}
  {(\nu^2k^4 + \omega^2)(\eta^2k^4 + \omega^2)}\
  \mathrm{d} k \mathrm{d}\omega.} 
  \label{10}
\end{eqnarray}
The expressions do not have definite signs so that it remains unclear  whether the large-scale Maxwell stress is increased or decreased by the turbulence. Even the  signs of $\kappa$ and $\kappa_\mathrm{p}$ may depend on the spectrum of the turbulence.

The simplest case is a turbulence with a {\em white-noise} spectrum containing all frequencies with the same amplitude.  Here and in the following we shall use  the Strouhal number ${\rm St}$ and the normalized characteristic frequency  ${w^*}$
\begin{equation}
 {\rm St}= \frac{u}{l_{\rm c}} \tau_{\rm c},  \quad\quad\quad {w^*}= \frac{w l_{\rm c}^2}{ \eta}
\label{S}
\end{equation}
($l_{\rm c}$ correlation length, $\tau_{\rm c}$ correlation time). 
The  turbulence frequency $w^*$ measures the characteristic frequency 
of the turbulence spectrum in relation to the diffusion frequency. 
It is large for flat spectra such as white noise and it is small for 
very steep spectra like $\delta$ functions. On the other hand, it is 
large in the high-conductivity limit and it is small in the 
low-conductivity limit. E.g., it is much larger than unity if 
the microscopic (Spitzer) diffusivity is used (high-conductivity limit). 
It should be unity if -- as it is used in large-eddy 
simulations -- $\eta=\eta_{\rm T}\simeq w l_{\rm c}^2$. 
In the numerical integrations presented below the 
limit $w^*\to  0$ (i.e. low-conductivity limit) applies to   
the case of the frequency spectrum  as a Dirac delta 
function $\delta(\omega)$.

The product of $\rm St$ and $w^*$ giving the magnetic Reynolds number 
\begin{equation}
 {\rm Rm}= \frac{u l_{\rm c}}{\eta},
\label{Rm}
\end{equation}
where we have used the relation $\tau_{\rm c}=1/w$ as a definition of the correlation time. Then it is $w^*= {\rm Rm/St}$.

In the high-conductivity limit (`white noise') one finds  the  simple results
\begin{equation}
\kappa= \frac{1}{15} {\rm Rm\ St}, \quad\ \ \ \ \ \ \ \ \ \ \kappa_{\rm p}= - \frac{1}{15} {\rm Rm\ St},
\label{kappa}
\end{equation}
so that the $\kappa$ is  positive and runs with 
${\rm Rm\ St}=u^2\tau_{\rm c}/\eta$ which is the (large) ratio of the eddy diffusivity 
and the microscopic diffusivity  hence  the magnetic tension is always (strongly) reduced. 
 On the other hand, the magnetic pressure is increased (see Eq.~\ref{4}). 
The  negative sign of the value of $\kappa_{\rm p}$  excludes the possibility that
 the effective pressure term in Eq.~(\ref{4}) changes its sign so that the total magnetic
  pressure becomes negative. This is formally possible after (\ref{kappa}) 
for the magnetic tension parameter $1-\kappa$.

The white-noise approximation, however,  is not perfect. If, for example, the opposite 
  frequency profile for very long correlation times, i.e. ${E \propto \delta(\omega)}$, 
is applied to (\ref{10}) then again the $\kappa$ is positive but the sign of $\kappa_{\rm p}$ 
depends on the magnetic Prandtl number
\begin{equation}
{\rm Pm}= \frac{\nu}{\eta}.
\label{PN}
\end{equation}
It is thus necessary to discuss the integrals in (\ref{10}) in more detail.
\subsection{  Pm $\vec{\geq 1}$}
For $\rm Pm>8$ one finds  that $\kappa_\mathrm{p}$ is negative-definite for {\em all}  possible spectral functions. The coefficient
$1-\kappa_\mathrm{p}$ of the magnetic pressure is thus positive-definite and cannot become negative.
This is not true for $\kappa$. We shall show that the $\kappa$ will `almost always' be positive so that the  Lorentz force term in the generalized Lorentz force expression (\ref{4b}) is `almost always' quenched by the existence of the turbulence.

For $\nu=\eta$ the expressions (\ref{10}) turn into
\begin{eqnarray}
\lefteqn{\kappa =\frac{1}{15} \int\limits_0^\infty\!\!\int\limits_0^\infty \frac{E k^2(3\eta^2 k^4-\omega^2)}{(\omega^2 + \eta^2 k^4)^2}\ {\rm d}k\,{\rm d}\omega,} \nonumber\\
\lefteqn{\kappa_{\rm p} = \frac{1}{15} \int\limits_0^\infty\!\!\int\limits_0^\infty \frac{E k^2(7\eta^2 k^4-9\omega^2)}{(\omega^2 + \eta^2 k^4)^2}\ {\rm d}k\,{\rm d}\omega ,}
\label{kap1}
\end{eqnarray}
which again do not have definite signs. One can write, however,  the expression for $\kappa$ as
\begin{eqnarray}
\lefteqn{\kappa=\frac{1}{15} \int\limits_0^\infty\!\!\int\limits_0^\infty
{\frac{2\eta^2 k^6 E}{(\omega^2+\eta^2 k^4)^2}}{\rm d}\, k\, {\rm d}\,\omega }\nonumber\\
&& \quad\quad\quad\quad\quad\quad -\ \frac{1}{15} \int\limits_0^\infty\!\!\int\limits_0^\infty{\frac{\omega k^2}{\omega^2+\eta^2 k^4}} \frac{\partial E}{\partial \omega} {\rm d}\, k\, {\rm d}\,\omega,
\label{kappp}
\end{eqnarray}
from which  $\kappa$ proves to be positive-definite for all spectral functions $E$ which do not increase  for increasing $\omega$. We shall see that the positivity of $\kappa$ which  reduces the effectivity of the angular momentum transport is a general result of the SOCA theory.

Further simplifications can be achieved by applying the model spectrum
\begin{equation}
  E(k,\omega) = q(k)\frac{2w}{\pi(w^2 + \omega^2)}
  \label{12}
\end{equation}
with
\begin{equation}
  \int\limits_0^\infty q(k)\ \mathrm{d} k = u^2,
  \label{12a}
\end{equation}
where $w$ is a characteristic frequency of the turbulence spectrum.  For $w\to 0$ (\ref{12}) represents a Dirac $\delta$-function while $w\to \infty$ gives `white noise'. The results are
\begin{equation}
\kappa=\frac{1}{15\eta}\int\limits_0^\infty \frac{w+3\eta k^2}{(w+\eta k^2)^2} q {\rm d} k 
\end{equation}
and
\begin{equation}
\kappa_{\rm p}=\frac{1}{15\eta}\int\limits_0^\infty \frac{7\eta k^2-w}{(w+\eta k^2)^2} q {\rm d} k .
\end{equation}
Again the  $\kappa$ is positive-definite. Note that for $w\to \infty$ the high-conductivity results (\ref{kappa}) are reproduced. In this case the $\kappa$'s are running with $1/\eta$ while for  $w\to 0$ the $\kappa$'s are running with $1/\eta^2$. This is a basic result: for low conductivity  and for high conductivity  the dependence of the $\kappa$'s on the magnetic Reynolds number $\rm Rm$ differs. For high conductivity  (white noise) the $\kappa$'s are proportionate to $\rm Rm$ while for low conductivity (steep spectra) the factor ${\rm Rm}^2$ appears. Note that for $\delta$-like spectral functions the numerical coefficient for $\kappa$ is 0.2 while for $\kappa_{\rm p}$ this factor is about 0.5. One can also find these values at the ordinate of Fig. \ref{f1}.

A basic difference  exists for $\kappa$ and $\kappa_{\rm p}$, too. While the $\kappa$ is positive-definite,  the $\kappa_{\rm p}$  can change its sign. Generally it will be positive only for small $w^*$  but it should be negative  for large $w^*$. Already from these arguments one finds the main complication of the problem. The shape of the turbulence spectrum has a fundamental meaning for the results.

To probe these results in  detail   a spectral function $q(k)$ 
\begin{equation}
q\simeq \frac{2 l_{\rm c}}{\pi} \frac{u^2}{1+k^2l_{\rm c}^2} 
\label{qu}
\end{equation}
is used. The  integration yields 
\begin{equation}
\kappa=\frac{1}{15} \frac{{\rm St\ Rm}\sqrt{w^*} (2+\sqrt{w^*})}{(1+\sqrt{w^*})^2},
\label{kappa0}
\end{equation}
so that
\begin{eqnarray}
\kappa\simeq\left\{
\begin{array}{c}
\frac{2}{15}{\rm Rm\ St}\sqrt{w^*}\\[4pt]
 \frac{1}{15} {\rm Rm\ St}
\end{array}
\right\} \quad {\rm for} \quad w^* \left\{\begin{array}{c} <1\\
\ >4\,.
\end{array}
\right.
\label{kappa1}
\end{eqnarray}
Hence, for small $w^*$ (low conductivity) the $\kappa$ runs with $\rm St^{0.5}\ Rm^{1.5}$
 while for high conductivity  the relation is simply $\rm St\ Rm$. One finds again 
 the differences between the two limits. In large-eddy simulations for the effective diffusivity
 the relation $\eta\simeq u l_{\rm c}$ is used so that $\rm Rm\simeq Pm\simeq 1$.
 As in the majority of the applications also the Strouhal number $\rm St$ is of the same 
 order the coefficient (\ref{kappa1}) is a small number. If for direct numerical simulations
 the numerical value of $\rm Rm$ becomes large then there is no reason that (\ref{kappa0})
 remains smaller than unity.

\begin{figure}  
\vskip-3mm
\hskip-7mm
  \includegraphics[width=8.95cm]{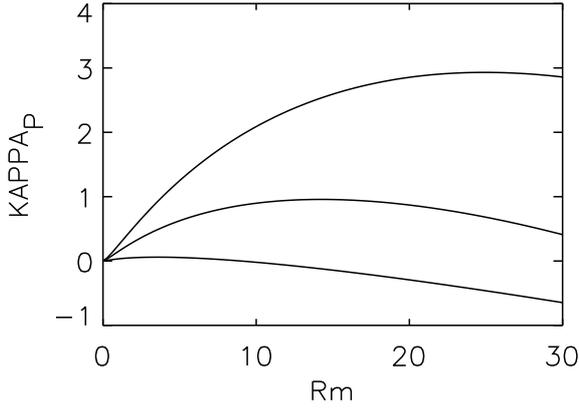}
\vskip-2mm
  \caption{The   $\kappa_{\rm p}$  vs. $\rm Rm$ after Eq. (\ref{kap2}). 
 \emph{From top to bottom}: $\rm St=7$, $\rm St=4$, and  $\rm St=1$. All curves have a 
  maximum. Note that all $\kappa_{\rm p}$ become negative for 
 sufficiently large $\rm Rm$; ${\rm Pm}=1.$
            }
	    \label{kappapressure}
\end{figure}

For  $\kappa_{\rm p}$  there is another situation. One   obtains
\begin{equation}
\kappa_{\rm p}=\frac{1}{15} \frac{{\rm Rm\ St} \ \sqrt{w^*} (3-\sqrt{w^*})}{(1+\sqrt{w^*})^2},
\label{kap2}
\end{equation}
hence,
\begin{eqnarray}
\kappa_{\rm p}\simeq\left\{
\begin{array}{c}
 \frac{1}{5}{\rm Rm^{1.5}\ St^{0.5}} \\[4pt]
 -\frac{1}{15} {\rm Rm\ St}\, 
\end{array}
\right\} \quad {\rm for} \quad w^* \left\{\begin{array}{c} <9\\
\ >9\, .
\end{array}
\right.
\label{kap4}
\end{eqnarray}
For $w^*>9$ the $\kappa_{\rm p}$ is  {\em negative} so that the total pressure is always positive. 
For $w^*<9$, however,   the $\kappa_{\rm p}$ becomes positive. In this case  for large Strouhal number 
    the total magnetic pressure ${1-\kappa_{\rm p}}$  becomes negative.  Figure  
\ref{kappapressure} demonstrates that $\kappa_{\rm p}$ exceeds unity for ${\rm St> 4}$. For ${\rm St> 4}$
one  finds ${\kappa_{\rm p}>1}$ for ${\rm Rm=14}$, i.e. ${w^*=3.5}$. We have to stress, however, 
that the SOCA approximation only holds if not both the quantities $\rm St$ and
 $\rm Rm$ simultaneously exceed unity. For turbulences in liquid metals in the
 MHD laboratory $\rm Rm\simeq 1$ is a typical value. Kemel et al. (2012) report
 an increase of $\kappa_{\rm p}$ with $\rm Rm^2$ (their Fig. 9, bottom). The 
 negative branch of (\ref{kap4})
 does not exist in the simulations ($\rm Pm=0.5$).

\subsection{Pm $\vec{\ll 1}$}\label{small}
The situation is more clear for   small magnetic Prandtl numbers,  which exist,
e.g., in stellar interiors, protoplanetary disks and also in the MHD laboratory.
It is possible to consider the limit $\nu\to 0$ in the Eqs. (\ref{10}) but only for   turbulence spectra with finite correlation time. Stationary patterns with $E\propto \delta(\omega)$ are excluded. In the limit of very small ${\rm Pm}$ the  Eqs.~(\ref{10}) reduce to
\begin{eqnarray}
\lefteqn{ \kappa =
  \frac{\pi}{15\eta}\int\limits_0^\infty E(k,0)\ \mathrm{d} k  -
  \frac{1}{15} \int\limits_0^\infty\!\!\int\limits_0^\infty
  \frac{E(k,\omega ) k^2}{\omega^2 +\eta^2k^4}
  \mathrm{d} k \mathrm{d}\omega, }
  \nonumber \\
\lefteqn{  \kappa_\mathrm{p} =
  \frac{4\pi}{15\eta}\int\limits_0^\infty E(k,0)\ \mathrm{d} k  -
  \frac{9}{15} \int\limits_0^\infty\!\!\int\limits_0^\infty
  \frac{E(k,\omega ) k^2}{\omega^2+\eta^2k^4}
  \mathrm{d} k \mathrm{d}\omega . }
  \label{11}
\end{eqnarray}
The spectrum (\ref{12})  leads to
\begin{eqnarray}
\lefteqn{\kappa = \frac{1}{15\eta w}\int\limits_0^\infty
  \frac{2\eta k^2 + w}{\eta k^2 + w}\ q(k)\ \mathrm{d} k ,}
  \nonumber \\
\lefteqn{  \kappa_\mathrm{p} = \frac{1}{15\eta w}\int\limits_0^\infty
  \frac{8\eta k^2 - w}{\eta k^2 + w}\ q(k)\ \mathrm{d} k .}
  \label{13}
\end{eqnarray}
Again  $\kappa$ is positive-definite. If the wave number spectrum has only a single value then 
\begin{eqnarray}
\lefteqn{\kappa = \frac{1}{15} \frac{{\rm Rm}^2}{w^*} \frac{2+w^*}{1+w^*},}
  \nonumber \\
\lefteqn{ \kappa_{\rm p}=\frac{1}{15}  \frac{{\rm Rm}^2}{w^*} \frac{8-{w^*}}{1+{w^*}}.} 
  \label{13}
\end{eqnarray}
The limit $w^*\to 0$ is here not allowed.  Again the  $\kappa_{\rm p}$   is positive (negative) for small (large) $w^*$. Formally, the  Strouhal number $\rm St$ does not appear. Replacing the $w^*$ by $\rm Rm/St$ in both limits the $|\kappa_{\rm p}|$ runs linearly with $ \rm St\ Rm$, i.e. with $1/\eta$.  

The  $\kappa$ also runs 
with $1/\eta$  in the high-conductivity limit, i.e.
\begin{equation}
\kappa\simeq \frac{1}{15}  {\rm St\ Rm}\, ,
\label{kappa2}
\end{equation}
while for low conductivity  the value is $
\kappa\simeq (2/15){\rm St\ Rm}$. For large  $w^*$, i.e. for $\rm Rm>St$, and
  $\rm Pm\la 1$ there is practically no influence of the numerical value of 
the magnetic Prandtl number (see Eq. \ref{kappa1}). Below we shall also demonstrate
 by numerical solutions of the integrals that Eq. (\ref{kappa2}) forms the main result 
of the present analysis. Whether the $\kappa$-coefficient may become larger
 than unity only depends  on the numerical values of $\rm St$  and $\rm Rm$. 
For large-eddy simulations with $\rm St=Rm=Pm=1$ the $\kappa$ is basically  only of order 0.1. 

With the spectral function (\ref{qu}) the results are very similar, i.e.
\begin{eqnarray}
 \kappa_{\rm p}=\frac{1}{15}  \frac{{\rm Rm}^2}{w^*} \frac{8-\sqrt{w^*}}{1+\sqrt{w^*}}. 
  \label{kappap}
\end{eqnarray}
This expression  only exceed unity for $\rm St\gg 1$.  For small $\rm St$ the sum $1-\kappa_{\rm p}$
 is thus always positive independent of the actual value of ${\rm Rm}$ contribution.

\begin{figure} 
\vskip-3mm
  \vbox{  
 \includegraphics[width=7.9cm]{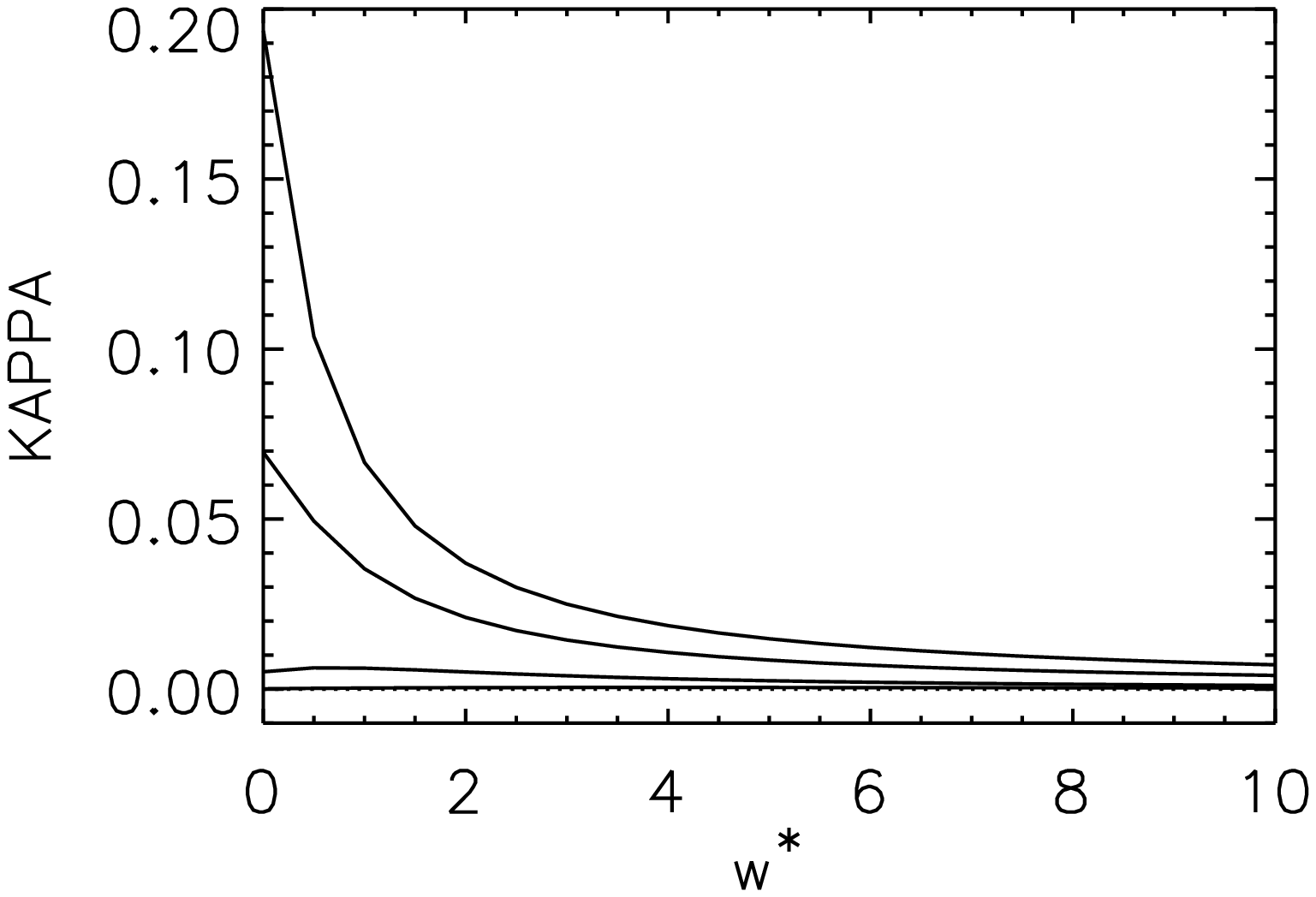}
\vskip-5mm
   \includegraphics[width=7.9cm]{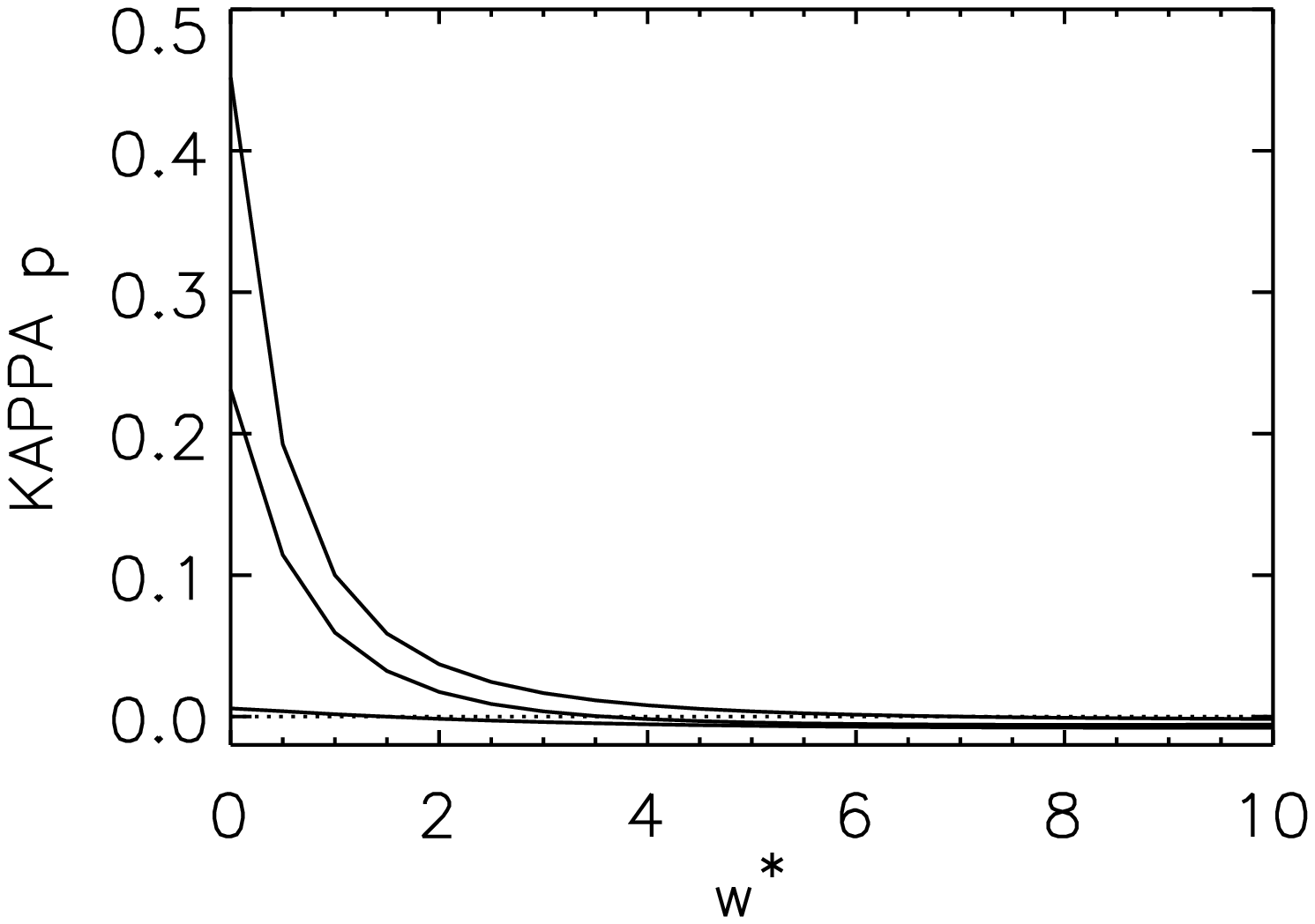}
\vskip-5mm
   \includegraphics[width=7.9cm]{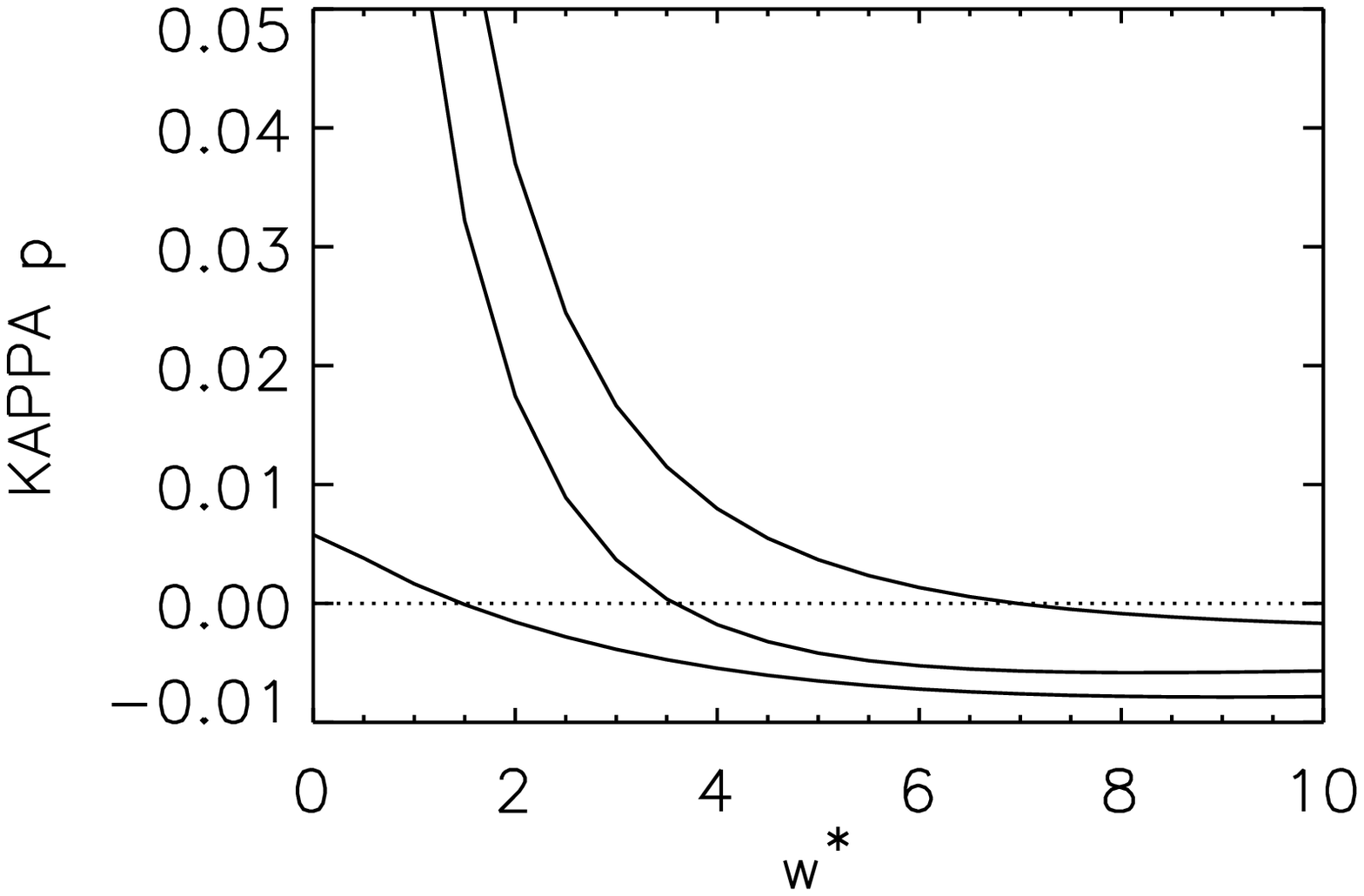}
}
\vskip-2mm
  \caption{The  $\kappa/{\rm Rm}^2$ (\emph{top}) and $\kappa_{\rm p}/{\rm Rm}^2$ 
(\emph{middle, bottom}) vs. $w^*$ for the one-mode model (\ref{14a}). 
The curves in the plots  (\emph{from top to bottom}) are for  $\rm S=0.01$, $\rm S=1$, $\rm S=3$,
 and $\rm S=10$.  At the left vertical axis the values are valid for the 
 delta function spectra (low-conductivity limit). 
 The quantities vanish as  $1/w^*$  for $w^*\to \infty$ 
 (high-conductivity limit, right vertical axis)
 leading to the result (\ref{kappa2}). \emph{Bottom}: details for $\kappa_{\rm p}/{\rm Rm}^2$;
 ${\rm Pm}=1$.
            }
	    \label{f1}
\end{figure}

\section{Strong fields}\label{strong}


\begin{figure}[htb]
\vskip-3mm
\hskip-5mm
  \vbox{  
  \includegraphics[width=8.9cm]{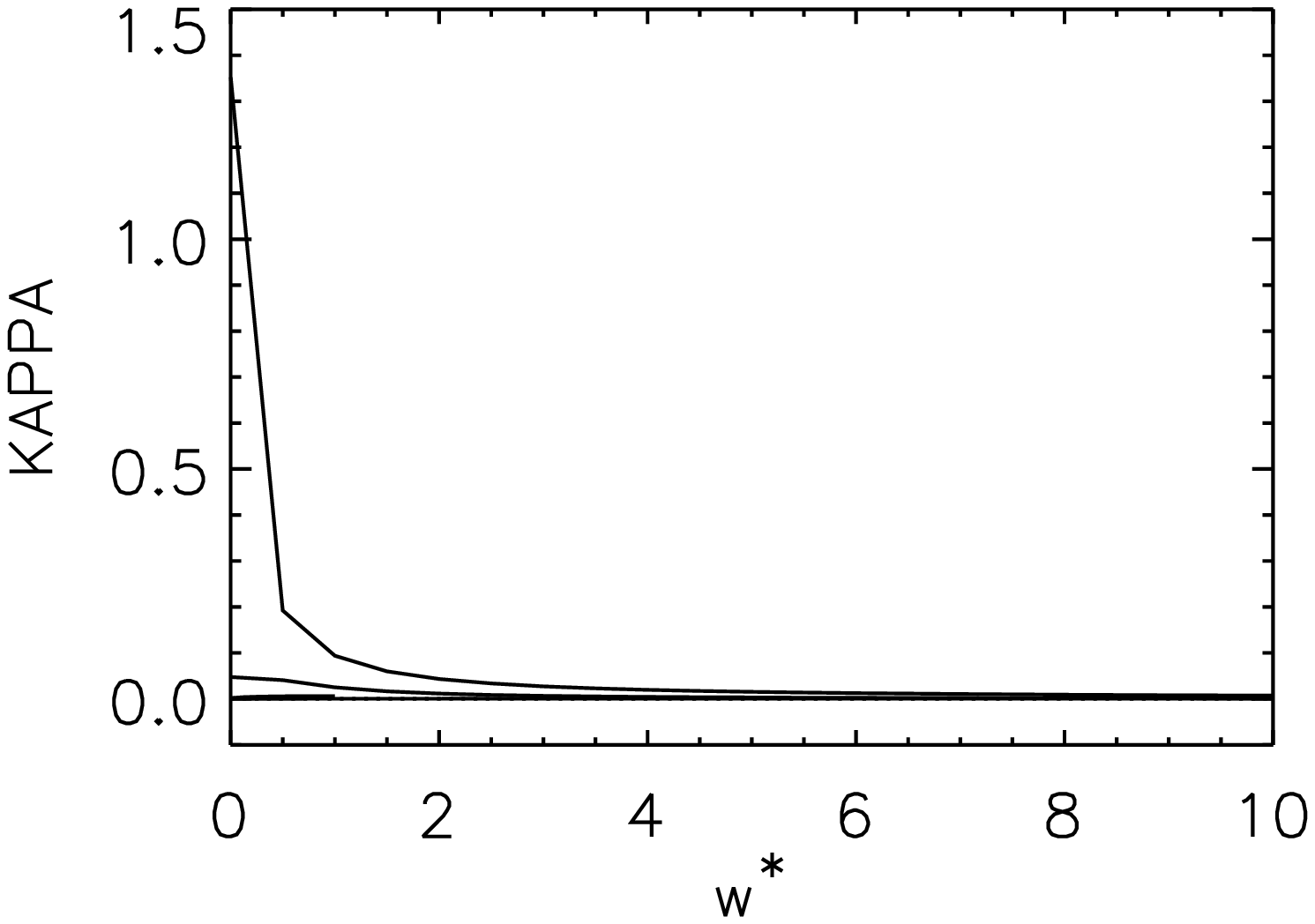}
\vskip-5mm
   \includegraphics[width=8.9cm]{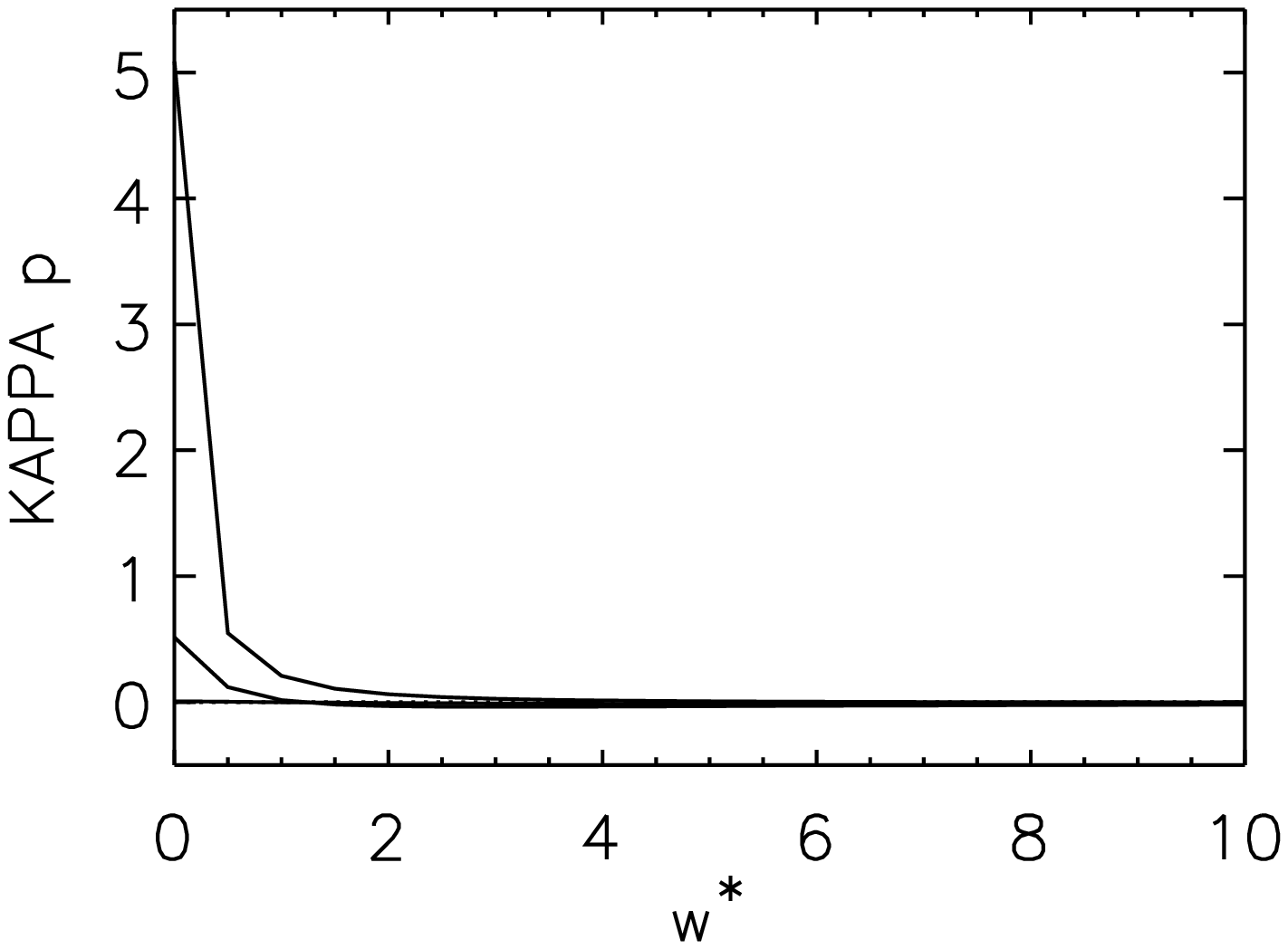}}
\vskip-2mm
  \caption{The  same as in Fig. \ref{f1} for ${\rm Pm}=0.1$.  
\emph{From top to bottom}: $\rm S=0.01$, $\rm S=1$, and $\rm S=10$.
 The quantities vanish as  $1/w^*$  for $w^*\to \infty$.
            }
	    \label{f2}
\end{figure}

So far only the influence of weak magnetic fields has  been
considered. The influence of strong magnetic fields is also
important to know.  The rather complex results of the SOCA theory with arbitrary magnetic
field amplitudes and with free values of both diffusivities are
given in the Appendix. These expressions can  be discussed by
applying the single-scale wave number spectrum
\begin{equation}
  q(k) = 2 u^2 \delta (k - l_{\rm c}^{-1}) 
  \label{14a}
\end{equation}
and the frequency spectrum (\ref{12}).
Such an  approximation allows to solve the Eqs. (A1)...(A4) numerically including the frequency integration so that the turbulence quantities $\kappa/{\rm Rm}^2$ and $\kappa_{\rm p}/{\rm Rm}^2$
only depend  on the Lundquist number  
\begin{equation}
{\rm S} = \frac{B l_{\rm c}}{\sqrt{\mu_0\rho} \eta}
\label{Lundquist}
\end{equation}
 of the  magnetic field, the frequency   $w^*$ and the magnetic Prandtl number $\rm Pm$. 

In the weak-field limit, ${\rm S \ll 1}$, one finds the overall result that $\kappa/{\rm Rm}^2$
 runs as $1/15 w^*$ (Figs.~\ref{f1} and \ref{f2}, top) so that again the general 
 result (\ref{kappa2}) is reproduced. For very small $w^*$, i.e. for
 delta function frequency spectra (or, what is the same, for very long correlation times), 
the $\kappa$'s run with $1/{\rm Rm}^2$ -- as already  shown above.

\begin{figure}
\vskip-3mm
\hskip-5mm
  \includegraphics[width=8.9cm]{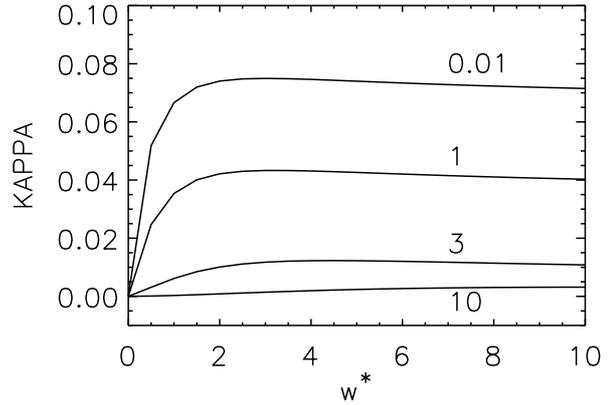}
\vskip-2mm
  \caption{The  verification of the relation (\ref{quench})  
 for the functions $w^*\kappa$ marked by their   Lundquist numbers $\rm S$. 
 The resulting value for  $\epsilon$ is about
   0.75; $\rm Pm=1$.
            }
	    \label{quenchplot}
\end{figure}

\begin{figure}
\vskip-3mm
\hskip-5mm
  \includegraphics[width=8.9cm]{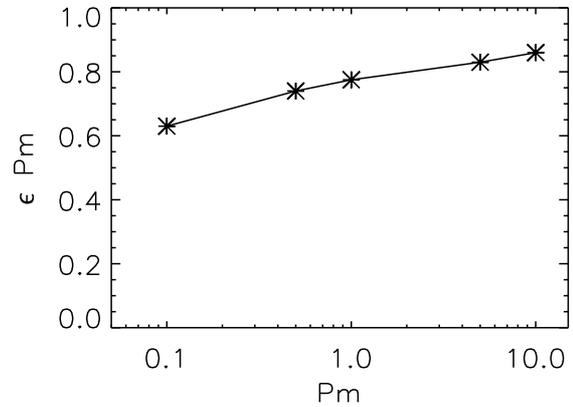}
\vskip-2mm
  \caption{The (weak) dependence of the quantity ${\rm Pm}\!\cdot \epsilon$ on the magnetic Prandtl number
 $\rm Pm$. }
	    \label{f3}
\end{figure}

When the field is not weak, the stress parameters  
rapidly decrease with $\rm S$. Figures  \ref{f1} and \ref{f2}   
also demonstrate that the magnetic quenching can be written as  
\begin{equation}
  \kappa\simeq \frac{\kappa_0}{1+\epsilon\  {\rm S}^2}
  \label{quench}
\end{equation}
(see Fig. \ref{quenchplot}), in confirmation to Brandenburg et al. (2010) 
who found the  
magnetic quenching in terms of $1/B^2$. From the  Figures one finds that  $\epsilon\lsim 1$ for $\rm Pm<1$. 
For large $\rm Pm$ the $\epsilon$ is even smaller. The magnetic quenching of 
the $\kappa$-parameter is thus  stronger for small 
 magnetic Prandtl number than for large $\rm Pm$. While a magnetic field with $\rm S=1$
 reduces the $\kappa$ remarkably if $\rm Pm<1$ in the opposite case  $\rm Pm>1$ 
the $\kappa$ is almost uninfluenced by $\rm S=1$. Figure \ref{f3} demonstrates the 
inverse dependence of the $\epsilon$ on the magnetic Prandtl number. One finds 
 ${\epsilon \simeq 0.75/{\rm Pm}}$. The quenching expression, therefore,  turns for $\rm Pm\neq 1$ into
\begin{equation}
  \kappa\simeq \frac{\kappa_0}{1+0.75\ {\rm Ha}^2}\, ,
  \label{quench}
\end{equation}
with the Hartmann number  ${\rm Ha= S/\sqrt{Pm}}$ instead of the 
Lundquist number $\rm S$. For the magnetic quenching it is thus not 
important which of the diffusivities is large and which is small. 
The quenching is very strong if one of them is small (see Roberts \& Soward
1975). 
For the high-conductivity limit ($\eta\to 0$) or for inviscid fluids  ($\nu\to 0$) the Hartmann  number $\rm Ha$ takes very large values so that even very  weak fields strongly suppress the $\kappa$-effect. 

Note that  
\begin{equation}
{\rm S}= {\rm Rm}\frac{ 
B}{B_{\rm eq}}\, ,
\end{equation}
 with ${B_{\rm eq}= \sqrt{\mu_0\rho\ \langle u^2\rangle} }$ as the equilibrium field strength. The magnetic quenching of the $\kappa$-term thus grows  with  $\rm Rm^2$ (Brandenburg \& Subramanian 2005) so that for growing $\rm Rm$ the $\kappa$  becomes smaller and smaller:
 \begin{equation}
  \kappa\simeq \frac{1}{15\epsilon} \frac{{\rm St}\ B_{\rm eq}^2}{{\rm Rm}\ B^2}.
  \label{quench1}
\end{equation}
It becomes  thus  clear that  in the high-conductivity limit even for rather small fields  the $\kappa$-term in Eq. (\ref{4}) takes very small values  which do not play an important role in the mean-field magnetohydrodynamics.

On the other hand, for $\rm Rm=1$  the well-known standard expression 
\begin{equation}
  \kappa= \frac{\kappa_0}{1+\epsilon\frac{  B^2}{B^2_{\rm eq}}}
  \label{quench1}
\end{equation}
for magnetic quenching appears with $\epsilon$ of order unity only slightly differing for small and large $w^*$.

Because of ${\rm St=\rm Rm=1}$ in this case the $\kappa$'s always 
remain smaller than unity in accordance to (\ref{kappa2}). Hence 
in both the possible concepts, i.e. the use of the microscopic 
diffusivities and the use of the  large-eddy simulations with 
subgrid diffusivities, the values of the turbulence-induced  
Maxwell tensor coefficients remain small.

The numerical simulations by Kemel et al. (2012) indeed yield a magnetic 
quenching of the {\em pressure} term in terms of $\rm Rm^2$ but only for $\rm Rm<10$.


\section{Catastrophic quenching?}
We have computed the  stress tensor which is formed by large-scale background fields, by the
 Reynolds stress of a turbulence field under the influence of the 
 field and the turbulent Maxwell stress of the field fluctuations. 
All contributions can be summarized in form of the classical Maxwell stress tensor
 but with  turbulence-modified  coefficients (see Eq.~\ref{4}).
 The modified  pressure term is now $1-\kappa_{\rm p}$ while the 
modified magnetic tension term is written as $1-\kappa$.
 The quantities $\kappa$ and $\kappa_{\rm p}$ have been computed within
 the quasilinear approximation (SOCA) which can be used if the minimum of
 both the numbers ${\rm St}$ and ${\rm Rm}$ is (much) smaller than unity.
 As almost all turbulences fulfill the condition ${{\rm St}\simeq 1}$,  the validity of SOCA requires
$
{{\rm Rm}= u l_c/\eta < 1.}
$
Under this restriction the resulting $\kappa'$s are always smaller than unity. 
For all magnetic Prandtl numbers ${\rm Pm}$ we found  $\kappa$ as positive so 
that the non-pressure force term $(\vec{B}\nabla)\vec{B}$ is reduced  under the
 influence of turbulence. This is in particular true for the coefficients of 
the angular momentum transport terms $B_\phi\, B_R$ and  $B_\phi\, B_z$ which, therefore,
  become more and more ineffective in turbulent fluids.

The sign of $\kappa_{\rm p}$ strongly depends on the magnetic Prandtl number. It proves to be negative-definite  for large ${\rm Pm}$. For smaller ${\rm Pm}$  the sign of $\kappa_{\rm p}$
depends on the shape of the frequency spectrum of the turbulence. For steep profiles, i.e. very long correlation times, the $\kappa_{\rm p}$ becomes positive while for flat  frequency-spectra of the turbulence which are as flat as the spectrum of white noise (very short correlation times) the  $\kappa_{\rm p}$ for ${\rm Rm}<1$ becomes  negative.

One could believe that relations valid for small ${\rm Rm}$ like
 ${\kappa_{\rm p}\propto {\rm St\!\cdot\! Rm}}$ can be also used for ${\rm Rm}>1$ so that 
finally the effective magnetic pressure becomes negative. This, however, is not true. 
The $\kappa_{\rm p}$ changes its sign for ${\rm Rm\gg St}$ and becomes negative. Hence,
 the total magnetic pressure results as  mostly positive. The only exception 
exists for sufficiently large $\rm St$ and sufficiently small $\rm Rm$ (see Fig. \ref{kappapressure}).

More dramatic is the situation with the magnetic tension and  its coefficient $1-\kappa$ which is also the coefficient of the vector $\vec{J}\times\vec{B}$ in the generalized Lorentz force in turbulent media. This coefficient is positive  for small $\kappa$, i.e. for sufficiently small $\rm Rm$ if $\rm St=1$. It  is  positive and smaller than unity  for the large-eddy simulations (`mixing-length model') considered at the end of  Sect. \ref{small} with $\rm Rm= St=Pm= 1$  (see Fig. \ref{f1}).

The question, however,  whether the $\kappa$ can 
exceed unity (so that $1-\kappa$ becomes negative) cannot 
finally be  answered within the quasilinear approximation. It  
is $\kappa \simeq 0.1\ \rm St\cdot Rm$ where one of the 
factors ${\rm St}$ and ${\rm Rm}$ {\em must} be smaller than 
unity but the product $\rm St\cdot Rm$ is formally { not} 
restricted by the SOCA. It is thus  a clear and surprising 
result also in the frame of SOCA that the angular momentum 
transport by large-scale magnetic fields  can strongly 
be  {\em suppressed}  under the influence of turbulence. 
The possible existence of an instability resulting from $\kappa>1$ has  
been confirmed 
by the numerical simulations by Brandenburg et al. (2011).

The formal background of this phenomenon  is that the 
integrals defining $\kappa$ and $\kappa_{\rm p}$ do not 
exist in the high-conductivity limit or, what is the same, 
in the ideal MHD. The same is true for the  much simpler 
magnetic-suppression problem of the eddy diffusivity. We 
take the expression
\begin{eqnarray}
\lefteqn{\eta_{\rm T} =\frac{1}{3} \int\limits_0^\infty \!\! \int\limits_0^\infty \frac{\eta k^2\ E}{\omega^2+\eta^2 k^4}}\nonumber\\
&& \quad\quad \left( 
1-\frac{6}{5} \frac{\eta^2 k^4-\omega^2}{(\omega^2+\eta^2 k^4)^2}\  \frac{\vec{B}^2}{\mu_0 \rho}
\right){\rm d}k\, {\rm d}\omega
\label{eta}
\end{eqnarray}
(Kitchatinov et al. 1994)
for the SOCA expression of the eddy diffusivity  under the presence of
a uniform magnetic  background field ($\rm Pm=1$).  The expression is part of a
series expansion which converges if the second term is smaller than
the first term.
The second term of the RHS of this expression  has two important properties:
 i) it is positive for all spectral functions $E$ with $\partial E/\partial \omega<0$
 so that the $\eta_{\rm  T}$ is always reduced by the magnetic fields,
 and ii) it does not exist for the limit $\eta\to 0$. In other words,
for rather small $\eta$ the integral becomes  large so that the magnetic quenching
 would be  extremely effective for large $\rm Rm$. This is why  such a 
 series expansion only holds for very weak fields. This  phenomenon has been 
called a `catastrophic' quenching (see Blackman \& Field 2000; Blackman \& Brandenburg 2002). 
It  exists within  the SOCA theory   for the eddy diffusivity and also for the eddy viscosity. 
One finds from Eq. (\ref{eta}) that the mentioned diffusivities are 
magnetically quenched like $1-{\rm S}^2$ for small $\rm S$ and like ${\rm S }^{-3}$ 
for large $\rm S$. Of course, by this procedure the $\eta_{\rm T}$ cannot become negative. 
We know, on the other hand, that the magnetic quenching of the eddy diffusivity in
 sunspots reduces its value (only) from $5{\times} 10^{12}$ cm$^2$\,s$^{-1}$ to about
 $10^{11}$ cm$^2$\,s$^{-1}$ what -- together with the time  decay law of the sunspots  -- can be
 understood with  quenching expressions like (\ref{eta})  for $\rm Rm=1$
 (R\"udiger \& Kitchatinov 2000). It is thus suggested to  work with the simple
 relations $\rm Rm=1$ and ${\rm S}\simeq B/B_{\rm eq}$ in applications  with turbulent convection.

Similarly, also the $\kappa$ increases for vanishing $\eta$. There is, however, 
no nonmagnetic term against which the magnetic influence can be neglected as it
 must be compared with the large-scale Lorentz force ${\vec{J}\times\vec{B}}$ 
 which is also of the second order in $\vec{B}$. The only possibility to keep
 the turbulence contribution  small for large $\rm Rm$ is to put ${\rm St\ll 1}$. 
However, if the magnetic field is super-equipartitioned then the $\kappa$ is
 magnetically quenched which introduces a new factor $\rm Rm^{-2}$. 
Then the magnetic-induced $\kappa$-effect finally  runs with $1/\rm Rm$ so that
 it vanishes in the high-conductivity limit. In summary, for large $\rm Rm$ and 
for very weak magnetic field the $\kappa$ can exceed unity (so that the stress tensor 
reverses sign) but this phenomenon  disappears already for rather weak fields.

\acknowledgements This work was supported by the Deutsche
Fo\-r\-schungs\-ge\-mein\-schaft and by the Russian Foundation for
Basic Research (projects 10-02-00148, 10-02-00391).

\appendix
\section{SOCA expressions of  the  $\vec{\kappa}$'s}
The  expressions for the mean-field Lorentz force  parameters $\kappa$ and $\kappa_\mathrm{p}$ of Eq.~(\ref{4}) provided by the quasilinear theory for arbitrary magnetic amplitudes can be written as 
\begin{eqnarray}
 \lefteqn{ \kappa = \int\limits_0^\infty \int\limits_0^\infty
  \frac{E(k,\omega ) k^2}{\omega^2+\eta^2 k^4}\ K(B,k,\omega )\
  \mathrm{d} k \mathrm{d}\omega ,} \nonumber\\
 \lefteqn{   \kappa_\mathrm{p} = \int\limits_0^\infty \int\limits_0^\infty
  \frac{E(k,\omega ) k^2}{\omega^2+\eta^2 k^4}\ K_\mathrm{p}(B,k,\omega )\
  \mathrm{d} k \mathrm{d}\omega .}
  \label{a1}
\end{eqnarray}
The  kernel functions $K$ and $K_\mathrm{p}$ depend on the magnetic field and the variables $k$ and $\omega$ via 
\begin{eqnarray}
\lefteqn{  \beta = \frac{k V}{\left(\omega^2+\eta^2k^4)^{1/4}
  (\omega^2+\nu^2k^4\right)^{1/4}} ,}\nonumber\\
  \lefteqn{
  LN = \log\left(\frac{\beta^2
  -2\beta\sin\frac{\phi}{2} + 1}{\beta^2 + 2\beta\sin\frac{\phi}{2} + 1}\right) ,}
  \nonumber \\
 \lefteqn{ AR = \arctan\left(\frac{\beta - \sin\frac{\phi}{2}}{\cos\frac{\phi}{2}}\right) + \arctan\left( \frac{\beta + \sin\frac{\phi}{2}}{\cos\frac{\phi}{2}}\right).}
  \label{a2}
\end{eqnarray}

Here, 
$
  \cos\phi = \left(\eta\nu k^4 - \omega^2\right)/\sqrt{(\omega^2+\eta^2k^4)
  (\nu^2 k^4 + \omega^2)}.
  $
The kernels read 
\begin{eqnarray}
\lefteqn{  K = \left(\frac{\omega^2+\eta^2k^4}{\omega^2+\nu^2k^4}\right)^{1/2}
  \frac{1}{8\beta^4}\left( - (\beta^2 + 3)\frac{LN}{4\beta\sin\frac{\phi}{2}}\ +\right.}\nonumber\\ 
\lefteqn{ \left. 
   (\beta^2 - 3)\frac{AR}{2\beta\cos\frac{\phi}{2}} \right) +
  \frac{1}{8\beta^4} \left( 6 - (\beta^2 -3 + 6\cos\phi )
  \frac{LN}{4\beta\sin\frac{\phi}{2}}\right. }\nonumber\\
 && \left. -\ (\beta^2 +3 + 6\cos\phi ) \frac{AR}{2\beta\cos\frac{\phi}{2}}\right)
  \label{a4}
\end{eqnarray}
and \newpage
\begin{eqnarray}
\lefteqn{  K_\mathrm{p} =  \left(\frac{\omega^2+\eta^2k^4}{\omega^2+\nu^2k^4}\right)^{1/2}
  \frac{1}{4\beta^4}\left( \frac{8}{3}\beta^2 + (\beta^2-1)\frac{LN}{4\beta\sin\frac{\phi}{2}}\ -\right. }\nonumber\\
  \lefteqn{ \left.(\beta^2 +1)\frac{AR}{2\beta\cos\frac{\phi}{2}}\right) +
  \frac{1}{4\beta^4} \left( 2 - (\beta^2 -1 + 2\cos\phi )
  \frac{LN}{4\beta\sin\frac{\phi}{2}} \right.}\nonumber\\
   && \left. -\ (\beta^2 + 1 + 2\cos\phi )
  \frac{AR}{2\beta\cos\frac{\phi}{2}}\right) .
  \label{a5}
\end{eqnarray}
The first parts in these expressions  represent  the contribution of 
the Reynolds stress while the  following lines represent the  small-scale Maxwell stress.
\end{document}